\documentclass[preprint]{aastex}

\shorttitle{Solar System Constraints} 
\shortauthors{Adams}

\newcommand{\be}{\begin{equation}} 
\newcommand{\ee}{\end{equation}}

\newcommand{\fsn}{ {\cal F}_{SN}}
\newcommand{\psn}{P_{>M}} 
\newcommand{\cross}{ {\langle \sigma \rangle}} 
\newcommand{\fuv}{ {\langle \langle F_{uv} \rangle \rangle}}
\newcommand{\fc}{ {\cal F}_C} 
\newcommand{\tcl}{ t_{\rm cl} } 
\newcommand{\pcl}{ P_{\rm cl} } 
\newcommand{\acl}{ A_{\rm cl} } 
\newcommand{\nexp}{ {\langle N \rangle}}

\begin{document}

\title{Constraints on the Birth Aggregate of the Solar System} 

\medskip 

\author{Fred C. Adams} 

\affil{Physics Department, University of Michigan, Ann Arbor, MI 48109} 

\email{fca@umich.edu} 

\bigskip 

\author{Gregory Laughlin} 

\affil{NASA Ames Research Center, Moffett Field, CA 94035} 

\email{gpl@acetylene.arc.nasa.gov} 

\begin{abstract} 

Using the observed properties of our solar system, in particular 
the isotopic compositions of meteorites and the regularity of the 
planetary orbits, we constrain the star formation environment of the
Sun within the scenario of (external) radioactive enrichment by a
massive star. This calculation yields a probability distribution for
the number of stars in the solar birth aggregate. The Sun is most
likely to have formed within a stellar group containing $N = \nexp
\approx 2000$ $\pm$ 1100 members.  The {\it a priori} probability of a
star forming in this type of environment is ${\cal P} \approx$ 0.0085,
i.e., only about 1 out of 120 solar systems are expected to form under
similar conditions. We discuss additional implications of this
scenario, including possible effects from the radiation fields
provided by the putative cluster environment and dynamical disruption
of the Kuiper Belt. The constraints of this paper place tight
restrictions on the properties of the solar birth aggregate for the
scenario of external enrichment by a massive star; alternately, these
tight constraints slightly favor a self-enrichment scenario for the
short-lived radioactive species.

\end{abstract}


\section{INTRODUCTION} 

A substantial fraction of star formation in our galaxy takes place in
groups or clusters, i.e., associations containing many young solar
systems.  As a consequence, our own solar system may have formed
within such a crowded environment. This hypothesis is bolstered by
observations of meteorites, which indicate that unexpectedly large
quantities of short-lived radioactive nuclei were present at the epoch
of planet formation. One traditional explanation for this set of
abundance anomalies is that the solar nebula was enriched in
radioactive species by a nearby massive star (see, e.g., the recent
review of Goswami and Vanhala 2000; Cameron et al. 1995; Cameron
1978).  Enrichment is usually envisioned to occur through a supernova
explosion, although Wolf-Rayet winds have also been invoked. In any
case, this scenario requires the presence of a nearby massive star,
which in turn implies that our solar system formed within a reasonably
large stellar group.  For completeness, we note that thermally pulsing
asymptotic giant branch stars have also been suggested as an
enrichment source (see Brusso, Gallino, and Wasserburg 1999), but the
probability that such a star is associated with a molecular cloud is
relatively low (Kastner and Myers 1994). 

We also know that our solar system could not have formed within a
cluster containing too many stellar members. In a sufficiently crowded
environment, the solar system would be disrupted by gravitational
scattering effects from passing stars, binary systems, and other solar
systems. The observed orbital elements of the outer planets exhibit
low eccentricities and are (almost) confined to the same orbital
plane. This relatively well-ordered configuration places modestly
tight restrictions on the characteristics of the solar birth
aggregate.  Some previous work has addressed this issue.  The
inclinations of the orbits of Uranus and Neptune are sensitive to
gravitational perturbations from passing stars in the solar birth
cluster; this effect implies a bound (Gaidos 1995; see also Tremaine
1991) on the product of the stellar number density and the residence
time in the cluster: $\int n dt < 3 \times 10^{4}$ pc$^{-3}$ Myr. In
this present work, we constrain the size of the solar birth aggregate
from both directions: It must be large enough to provide a
sufficiently massive star for enrichment and small enough to allow 
for well-ordered planetary orbits. 

We must keep in mind, however, another possible way to explain the
inferred abundances of radioactive species: The solar nebula can be
self-enriched through energetic processes in its early formative
stages (e.g., Lee et al. 1998). This latter scenario allows for the
solar system to form in relative isolation.  Recent evidence (McKeegan
et al. 2000) provides support for the self-enrichment scenario. This
work indicates the presence of $^{10}$Be, a short-lived radioactive
species that cannot be produced in supernovae and hence must be
produced by nuclear reactions from energetic particles (as would be
expected in a picture of self-enrichment).  An important issue for
solar system formation is to decide between these two alternate
enrichment scenarios. As we show here, the external enrichment
scenario places strong constraints on the birth environment of the
solar system. These tight constraints, in turn, imply a low
probability of success for the external enrichment scenario and 
thus suggest that self-enrichment may be more viable. 

In this paper, we explore the compromise necessary to have the solar
system formed within a stellar group large enough to produce a nearby
massive star and, simultaneously, small enough to allow the planetary
orbits to remain relatively unperturbed. We first determine the
probability of a stellar cluster containing a sufficiently massive
star as a function of the number $N$ of cluster members (\S 2.1). We
then calculate the cross sections for passing stars to disrupt the
orbits of the outer planets in the solar system (\S 2.2), and use the
results to estimate the probability of disruption as a function of
cluster size $N$ (\S 2.3).  Folding these results together, we find
the probability distribution for the size $N$ of the solar system's
birth aggregate (\S 2.4), the corresponding expectation value $\langle
N \rangle$ for the cluster size, and the {\it a priori} probability of
a star forming in such an environment (\S 2.5). Next, we use these
results to reconstruct the expected ultraviolet radiation field
impinging upon the solar nebula during the epoch of planet formation
(\S 2.6). Finally, we calculate the cross sections for the scattering
of Kuiper Belt objects through gravitational interactions with passing
stars (\S 2.7) and thereby place further constraints on the birth
aggregate. We conclude (\S 3) with a summary and discussion of our
results.

\section{CONSTRAINTS ON THE SOLAR BIRTH ENVIRONMENT} 

\subsection{Isotopic Abundances and the Probability of a Massive Star} 

The observed isotopic abundances in solar system bodies provide strong
constraints on the material originally contained within the solar
nebula. In particular, isotopic studies of meteorites indicate that
the solar nebula was enriched in many short-lived nuclides while the
solar system was still forming (for a recent review, see Goswami and
Vanhala 2000). One leading explanation for this observed enrichment is
that a nearby supernova explosion detonated during the formative
stages of the solar system and thereby enriched the solar nebula.
Possible enriching stellar sources include not only a supernova by a
massive star (with mass $M_\ast > 25 M_\odot$; Cameron et al. 1995),
but also a non-exploding Wolf-Rayet star (with mass $M_\ast > 60
M_\odot$; Arnould et al. 1997). In either case, however, the Sun would
have to be born in close proximity to a massive star. The required
distance is $\sim2$ pc (Goswami and Vanhala 2000), so the Sun would
have to be contained with the same cluster as the massive star, but
would not require a particularly special location within the
cluster. As mentioned earlier, however, this massive star solution is
not the only one: Self-enrichment by cosmic rays produced by the early
Sun itself remains a possibility (Lee et al. 1998), and the
constraints derived in this paper may ultimately help distinguish
between these two competing scenarios.

For this work, we want to calculate the probability that the birth 
aggregate of the Sun contained a sufficiently massive star (to either
provide the supernova or Wolf-Rayet effects). And we need to know this
probability as a function of the number $N$ of stars in the birth
aggregate. The probability $\psn$ of a group of stars having a massive
star, say with mass greater than $M$, is equal to $1 - P_{\rm not}$, 
where $P_{\rm not}$ is the probability of the group {\it not} having 
a massive star. We choose this approach because the probability of 
not having a massive star is straightforward to calculate. Let $p_C$
be the probability that a star is {\it not} larger than some
pre-specified mass scale $M_C$. The probability of a star (a group
containing only one star) not having a massive star is thus $p_C$. The
probability for a group of $N$ stars not having a massive star is thus
$P_{\rm not}$ = $p_C^N$. The probability $\psn$ of the group containing 
a massive star is then given by the expression 
\be
\psn = 1 - p_C^N \, . 
\label{eq:psuper} 
\ee

The probability $p_C$ is determined by the stellar initial mass
function (IMF). Since only the most massive stars enter into this
present application, we need to specify the high mass tail of the IMF
(with an appropriate normalization). We let $\fsn$ denote the fraction
of stars that are large enough ($M_\ast > M_{SN} \approx 8 M_\odot$)
to explode as supernovae at the end of their nuclear burning lives;
for a standard IMF, $\fsn \approx 0.003$.  The observed solar system
enrichment requires larger stars with mass, say, $M_C$ = 25 -- 60
$M_\odot$.  For a given mass requirement $M_C$, the fraction of stars
that are heavy enough to provide enrichment is then given by the
expression $\fc$ = $\fsn (M_{SN}/M_C)^\gamma$, where $\gamma$ is the
power-law index of the IMF for high masses [i.e., $df / d \ln m$
$\sim$ $m^{-\gamma}$, where $\gamma$ = 1.35 is the traditional
Salpeter (1955) value].  Observations of high mass stars in rich
clusters (e.g., Brandl et al. 1996) indicate slightly larger indices
for the high end of the IMF, $\gamma \sim 1.6$, although the values
show some variation from cluster to cluster. Using $\gamma$ = 1.6 and
the required mass scale $M_C$ = 25 $M_\odot$, we obtain $\fc$ = 4.85
$\times 10^{-4}$, which implies $p_C$ = $1 - \fc$ = 0.999515.  We
adopt this value for the remainder of the paper.  To include other
possible parameter choices, we can immediately generalize equation
[\ref{eq:psuper}] to obtain 
\be
\psn = 1 - \Bigl[ 1 - \fsn (M_{SN}/M)^\gamma \Bigr]^N \, . 
\ee
For our standard choice of parameters, we can directly find the number
of stars required for a cluster to have a 50-50 chance of containing a
sufficiently massive star: $N = - \ln2 / \ln p_C$ $\approx$ 1430.

\subsection{Cross Sections for Orbital Disruption} 
 
We want to ultimately calculate the probability that a solar system
will be disrupted as a function of the number $N$ of stars in its
birth aggregate. The disruption rate of a solar system is given by 
\be 
\Gamma = n \sigma v \, , 
\ee
where $\sigma$ is the disruption cross section, $n$ is the mean
density of other systems, and $v$ is the relative velocity (typically,
$v \sim 1$ km/s). 

Using our planet scattering code developed previously (Laughlin and
Adams 1998, 2000), we can calculate the cross sections for the
disruption of our solar system. In particular, we want to find the
effective cross section $\cross$ for a specified change in orbital
parameters resulting from scattering encounters with other cluster
members (which are mostly binaries).  Although we should, in
principle, consider all possible encounters, no matter how distant,
only sufficiently close encounters have an appreciable contribution 
to the cross section.  We thus define our effective cross sections 
through the relation 
\be
\cross \equiv \int_{0}^{\infty} f_D (a) (B \pi a^{2}) p(a) \, da \, , 
\label{eq:netcross} 
\ee
where $a$ is the semi-major axis of the binary orbit and $p(a)$
specifies the probability of encountering a binary system with a given
value of $a$.  For a given value of $a$, we thus include only those
scattering interactions within the predetermined area $B \pi a^{2}$,
where $B$ is a dimensionless factor of order unity.  The function
$f_D (a)$ specifies the fraction of encounters which result in a
particular outcome (for scattering between the Solar System and a
binary of semi-major axis $a$).  Because we neglect the contribution
to the cross section from scattering interactions outside the area $B
\pi a^{2}$, equation [\ref{eq:netcross}] provides a lower limit to the
true cross section.
 
The distribution $p(a)$ is determined by the observed distribution 
of binary periods and by the normalization condition 
\be 
\int_{0}^{\infty} p(a) da = 1 \, .
\ee 
We model the observed period distribution, and hence obtain $p(a)$, 
by fitting the results of Kroupa (1995).  The observed binary period 
distribution peaks at $P$ = $10^5$ days, but the distribution is 
relatively broad and significant numbers of binaries have periods
longer than $10^{7}$ days.  For this set of scattering experiments, 
however, we only include binaries with $a < 1000$ AU because binaries
with larger values of $a$ have little contribution to the cross
sections.
 
The set of possible encounters which can occur between the solar
system and a field binary is described by 10 basic input parameters.
These variables include the binary semi-major axis $a$, the stellar
masses, $m_{\ast 1}$ and $m_{\ast 2}$, of the binary pair, the
eccentricity $\epsilon_{\rm b}$ and the initial phase angle 
$\ell_{\rm b}$ of the binary orbit, the asymptotic incoming velocity
$v_{\rm inf}$ of the solar system with respect to the center of mass
of the binary, the angles $\theta$, $\psi$, and $\phi$ which describe
the impact direction and orientation, and finally the impact parameter
$h$ of the collision.  Additional (intrinsic) parameters are required
to specify the angular momentum vector and initial orbital phases of
the planets within the solar system.

To compute the fraction of disruptive encounters $f_D (a)$ and hence
the corresponding cross sections, we perform a large number of
separate scattering experiments using a Monte Carlo scheme to select
the input parameters.  Individual encounters are treated as $N$-body
problems in which the equations of motion are integrated using a
Bulirsch-Stoer scheme (Press et al. 1986).  During each encounter, we
require that overall energy conservation be maintained to an accuracy
of at least one part in $10^{8}$. For most experiments, both energy
and angular momentum are conserved to better than one part in
$10^{10}$.  This high level of accuracy is needed because we are
interested in the resulting planetary orbits, which carry only a small
fraction of the total angular momentum and orbital energy of the
entire system (for further detail, see Laughlin and Adams 1998, 2000).
 
For each scattering experiment, the initial conditions are drawn from
the appropriate parameter distributions. More specifically, the binary
eccentricities are sampled from the observed distribution (Duquennoy
and Mayor 1991).  The masses of the two binary components are drawn
separately from a log-normal initial mass function (IMF) which is
consistent with the observed distribution of stellar masses (as
advocated by Adams and Fatuzzo 1996).  For both the primary and the
secondary, we enforce a lower mass limit of 0.075 $M_\odot$ and hence
our computed scattering results do not include brown dwarfs.
Observational surveys indicate that brown dwarf companions are
intrinsically rare (Henry 1991); in addition, this cutoff has only a
small effect because our assumed IMF peaks in the stellar regime. The
impact velocities at infinite separation, $v_{\rm inf}$, are sampled
from a Maxwellian distribution with dispersion $\sigma_{v}$ = 1
km/sec, which is a typical value for stellar clusters (Binney and
Tremaine 1987). The initial impact parameters $h$ are chosen randomly
within a circle of radius $2a$ centered on the binary center of mass
(using a circular target of radius $2a$ implies that $B$=4 in equation
[\ref{eq:netcross}]).
 
Using the Monte Carlo technique outlined above, we have performed
$N_{\rm exp}$ $\approx$ 50,000 scattering experiments for collisions
between binary star systems and the outer solar system.  These 7-body
interactions involve all four giant planets, the Sun, and the two
binary members.  From the results of these experiments, we compute the
cross sections for orbital disruption of each outer planet (according 
to equation [\ref{eq:netcross}]). 

The cross sections for the giant planets to increase their orbital
eccentricities are shown in Table 1.  For each given value of
eccentricity $\epsilon$, the entries give the cross sections [in units
of (AU)$^2$] for the eccentricity to increase to any value greater
than the given $\epsilon$; these cross sections include events leading
to either ejection of the planet or capture by another star. The
listings for $\epsilon$ = 1 thus give the total cross sections for
planetary escape and capture (taken together). In the two additional
lines below the main part of the table, we also present the cross
sections for planetary escapes and captures separately.  For each
cross section listed in Table 1, we also provide the one standard
deviation error estimate for the Monte Carlo integral; this quantity
provides a rough indication of the errors due to the statistical
sampling process (Press et al. 1986).  In this work, we are mostly
interested in the largest possible cross sections for disruption,
which ultimately provide the tightest constraints on the environment
of the early solar system.  Of the four giant planets, Neptune is the
most easily sent into an alternative orbits, as expected (see Table
1).  For the disruption of Neptune, we find a cross section $\cross
\approx$ 143,000 AU$^2$ to increase its eccentricity to $\epsilon >$
0.1 and $\cross \approx$ 167,000 AU$^2$ to double its eccentricity.
For this work, we use the cross section for Neptune to double its
orbital eccentricity [about (400 AU)$^2$] to represent the effective
cross section for solar system disruption (through eccentricity
increases).

Another way for the solar system to be disrupted is by changing the
planes of the planetary orbits. Table 2 shows the cross sections for
the inclination angles of the planetary orbits to increase by varying
amounts. The scattering experiments start with all four giant planets
in the same plane. After each collision, the angular momentum vectors
of the planets will not, in general, be aligned. The quantity $\Delta
\theta_i$ in the table is the largest angle (given in radians) between
any two of the angular momentum vectors for the four planets.  The
cross section for any one of the planets to escape or be captured is
$\cross \approx 17100 \pm 420$ (in units of AU$^2$).  The second
column of Table 2 gives the cross sections for all four planets to
remain bound to the Sun, but have at least one of the relative angles 
exceed $\Delta \theta_i$.  The final column gives the total disruption
cross sections, including planet escapes, planet captures, and the
inclination angle increases.  Notice that the cross section for
planetary escape and/or capture is larger than the cross section for
scattering events to increase the inclination angles beyond $\Delta
\theta_i \approx 2.4 \approx 3 \pi/4$.  As is well known (Shu 1980),
the inclination angles for the (present-day) planetary orbits in our
solar system show a small spread, only about 3.5 degrees or 0.061
radians.  The cross section for the inclination angles to increase to
this maximum allowed value is about $\cross \approx$ 158,000 AU$^2$,
which is comparable to the cross section for the maximum allowed
eccentricity increase determined earlier.

To summarize, our solar system can be disrupted within its birth
aggregate by scattering events which increase both the eccentricities
and the inclination angles of the orbits of the outer planets. But
these two effects are {\it not} independent -- scattering events that
pump up the orbital eccentricity also increase the inclination angles
-- so we cannot add the cross sections. As a rough benchmark, the
total cross section for significant disruption (large enough to be
inconsistent with observations of the present day solar system) is
thus $\cross \approx$ 160,000 AU$^2$ = (400 AU)$^2$. We use this
cross section as a representative value for the rest of this paper.

\subsection{Cluster Evolution and the Probability of Solar System Disruption}

In order to assess the likelihood of planetary disruption, we must 
fold into the calculation considerations of the background cluster
environment.  The effective ``optical depth'' $\tau$ to disruption 
can be written in the form 
\be
\tau = \int \Gamma dt \, = \cross \int_0^{\tcl} v n dt \, , 
\label{eq:taudef0} 
\ee
where the cross section $\cross$ is now considered as a known quantity. 
The integral is taken over the total lifetime $\tcl$ of the cluster
environment. The velocity scale $v$, the mean stellar density $n$, and
the lifetime $\tcl$ are (on average) increasing functions of the
number $N$ of stars in the system. We are implicitly assuming that the
Sun stays in the birth cluster for most of the cluster lifetime. This
assumption is reasonable because the Sun is relatively heavy and will
tend to sink toward the cluster center rather than become immediately
evaporated. In addition, the Sun {\it must} stay in the cluster long
enough for massive stars to evolve and die (in order to have the solar
system enriched in short-lived radioactive species).  For purposes of
this paper, the integral can be approximated as follows: 
\be
\int_0^{\tcl} v n dt \,  \approx 50 n_0 v_0 t_{R0} \, , 
\label{eq:t100} 
\ee 
where the subscripts indicate that the quantities are evaluated at
their initial values and where $t_{R0}$ is the initial relaxation
time. In making the approximation [\ref{eq:t100}], we have used the 
result that the typical lifetime of a cluster is about 100 times the
initial dynamical relaxation time $t_{R0}$ (see Binney and Tremaine
1987), but the number density of other stellar systems (both singles 
and binaries) is less than the starting value (averaged over $\tcl$). 
Since $n_0 \approx N/4 R^3$ and $t_{R}$ $\approx$ $(R/v) N/(10
\ln N)$, we find that $n_0 v_0 t_{R0}$ $\sim$ $N^\mu$, where the index
$\mu \approx 2$.  Assuming this power-law scaling for the
$N$-dependence, we can ``evaluate'' the optical depth integral and
write it in the form 
\be
\tau = (N/N_C)^\mu \, , 
\label{eq:taudef} 
\ee
where $N_C$ is the number of stars in the system required to make the 
optical depth unity. Throughout this work, both $N$ and $N_C$ refer to 
the number of singles and binaries, where the binaries are counted as 
one other stellar system. In other words, $N$ and $N_C$ represent the 
number of primaries. For our adopted standard value $\cross$ = 
(400 AU)$^2$ $\approx 4 \times 10^{-6}$ pc$^2$, equation 
[\ref{eq:t100}] implies that 
\be
{N_C \over \sqrt{\ln N_C} } \approx \bigl( 1.25 \cross \bigr)^{-1/2} 
R \approx 447 \Bigl( {R \over 1 {\rm pc}} \Bigr) \, . 
\label{eq:nc}
\ee
For the representative value $R$ = 2 pc, we find $N_C \approx 2500$, 
which we take as our standard value for the remainder of this 
paper. For observational comparison, the Trapezium cluster in Orion
has approximately 2300 stars in its central region of size $R \sim 2$
pc (Hillenbrand and Hartmann 1998); if the cluster lives for 100 initial 
relaxation times $t_{R0}$ (again, see Binney and Tremaine 1987), where 
$t_{R0}$ $\approx$ 15 Myr, then the effective optical depth for the 
Trapezium cluster will be about $\tau \approx$ 0.9, in reasonable 
agreement with the choices of parameters taken here. 

The probability $P(t)$ of the solar system surviving (i.e., not being 
disrupted) is given by the solution to the simple differential equation 
\be
{dP \over dt} = - \Gamma P \qquad \to \qquad 
P(t) = \exp \bigl[ - \int^t \Gamma dt \bigr] \, ,
\ee
where we have used the fact that $P(t=0)$ = 1.  Putting the above
results together, we obtain the probability $P_{\rm dis}$ for the
solar system surviving disruption as a function of the cluster 
size $N$, i.e., 
\be
P_{\rm dis} (N) = \exp \bigl[ - (N/N_C)^\mu \bigr] \, ,
\ee
where we expect $\mu = 2$ and $N_C \approx 2500$. 

\subsection{Probability Distribution for the Size of the 
Solar Birth Aggregate} 

By considering both constraints derived above, i.e., by assuming that
the early solar system experienced a supernova and that the outer
planets were not severely disrupted, we obtain the probability distribution
that the Sun formed in a birth aggregate containing $N$ members. This
probability distribution $P_\odot$ is given by  
\be
P_\odot (N) = \psn P_{\rm dis} = 
\bigl (1 - p_C^N \bigr) \, \exp \bigl[ - (N/N_C)^\mu \bigr] \, . 
\label{eq:joint} 
\ee 
The resulting joint probability distribution is shown in Figure 1.
The peak of the distribution occurs at $N \approx 1465$. However, the
more relevant quantity is the expectation value, $\nexp \equiv \int N
P_\odot (N) dN$ / $\int P_\odot dN$, which is somewhat larger, $\nexp
\approx$ 1970.  We also obtain a measure of the range of allowed
cluster sizes:  The variance of the distribution $\langle (\Delta N)^2
\rangle^{1/2}$ = 1090, so we obtain $N \approx 2000 \pm 1100$.
Alternately, the allowed range can be defined by the half-maximum 
points of the probability distribution, i.e., 425 $< N <$ 3000. 

To further illustrate the conditions within the required birth
aggregate, let's now consider the evolution of a stellar system with
$N=2000$ stars. This number of stars should be calculated {\it after}
gas removal from the cluster. In the initial stages of cluster
formation, the system will contain some fraction of its mass in
gaseous form as well as additional stars. The stars obey a
distribution of velocities. After gas removal, the high velocity stars
leave the system and the low velocity stars remain behind. As a
benchmark, if the star formation efficiency is 50\%, and the initial 
velocity distribution of the stars is isotropic, then the cluster must
initially contain about 2560 stars (Adams 2000). This value implies that 
the solar birth aggregate is comparable to (but somewhat larger than) 
the present day Trapezium cluster in Orion (which has about 2300 stars 
within its central 2 pc; see Hillenbrand and Hartmann 1998). 

After gas removal, our benchmark cluster contains 2000 stars and has a
total mass of about 1400 $M_\odot$. With a typical radial size of $R =
1.5$ pc, the velocity scale will be about 2 km/s.  The dynamical
relaxation time of the system is initially $t_{R0}\sim$20 Myr, whereas
the crossing time is only $t_{\rm cross}\sim$0.75 Myr. The
characteristic time scale $t_P$ for nebular disks to retain their gas
and form giant gaseous planets is $\sim$10 Myr (e.g., Lissauer 1993),
which is roughly comparable to (but shorter than) the dynamical
relaxation time $t_R$, but much longer than the crossing time $t_{\rm
cross}$ of the system. All of these time scales are much less than the
total cluster lifetime $\tcl$, which is, in turn, much shorter than
the current age of the solar system $t_\odot \approx$ 4.6 Gyr.  The
time scales involved in the problem thus obey the ordering 
\be
t_{\rm cross} \ll t_P < t_R \ll \tcl \ll t_\odot \, , 
\label{eq:timeorder} 
\ee
which determines the timing of relevant events. The solar system will
experience many orbits through the birth cluster while its planets
form, but the background gravitational potential and structure of the
cluster will not change appreciably. The solar system will thus
randomly sample the cluster volume during the planet forming epoch.
Since the planet formation time is much shorter than the total cluster
lifetime, $t_P \ll \tcl$, the planets are available for disruption for
most of the cluster's life (as implicitly assumed above). Furthermore,
because the planets have a relatively small (but significant) chance
of being disrupted during the entire life span of the cluster, they
have a much smaller chance of being disrupted during their early
formative stages (smaller by a factor of $t_P/\tcl$ $\ll$ 1).  In
general, if the cluster is sufficiently diffuse to allow the planetary
orbits to remain unperturbed ({\it after} planet formation), then the
cluster environment can have relatively little effect on the planet 
formation process (via dynamical scattering processes; radiation 
can play an important role as discussed in \S 2.6). 

The initial stellar density of the cluster is about 200 pc$^{-3}$ when
averaged over the whole system. As the cluster evolves, the half-mass
radius of the cluster remains roughly constant, with the central
regions shrinking inwards and the outer regions expanding (Binney and
Tremaine 1987).  The average density over the cluster lifetime (about
100 initial relaxation times or $\sim$1 Gyr), is thus $n \sim 50 - 100$
pc$^{-3}$. The quantity $\int n dt \sim 10^5$ pc$^{-3}$ Myr, which is
comparable to a previous constraint on this quantity (derived in
Gaidos 1995).  Not surprisingly, the expected optical depth to
significant scattering events, $\tau = \cross v \int n dt \approx$ 0.4
$\langle v \rangle$ (1 km/s)$^{-1}$, implies that solar system has a
reasonable but not overwhelming probability of disruption. In order
for this value of the optical depth to agree with our approximation of
equation [\ref{eq:taudef}], we would need an average velocity
dispersion of $\langle v \rangle \approx$ 1.6 km/s (again, a
reasonable value).

We can generalize this calculation to include other choices of
parameters. One quantity that is not well determined is the number of
stars $N_C$ required for a cluster to have optical depth unity for
solar system disruption (see equations [\ref{eq:taudef0} --
\ref{eq:nc}]). This parameter $N_C$ ultimately depends on the degree
of cluster concentration and the immediate galactic environment, which
show some variation and are not completely well known. We also might
want to consider other choices for the mass scale $M_C$ of the
required massive star (e.g., the Wolf-Rayet scenario requires a larger
value) or different choices of the IMF, both of which lead to
different values of $p_C$. Fortunately, however, we can analytically
determine the expectation value $\nexp$ for the size of the solar
birth aggregate (for the standard choice of $\mu=2$ in equation
[\ref{eq:joint}]). We find 
\be 
\nexp = \beta N_C {{\rm e}^{\beta^2} \bigl[ 1 - {\rm Erf} (\beta) \bigr] 
\over 1 - {\rm e}^{\beta^2} \bigl[ 1 - {\rm Erf} (\beta) \bigr] } \, , 
\label{eq:ngeneral} 
\ee
where Erf$(x)$ is the error function and where we have defined a
dimensionless parameter $\beta \equiv 0.5 N_C (- \ln p_C)$.  Equation
[\ref{eq:ngeneral}] thus provides the expectation value for the size
of the solar birth aggregate for any value of $N_C$ and/or $p_C$.

\subsection{Probability of a Star Forming under Solar Conditions} 

Next, we can find the overall probability that a star forms in an
environmental group containing $N = \nexp \approx 2000$ members by
using the probability distribution $P_\odot$ given in equation
[\ref{eq:joint}].  This calculation gives us a feeling for how common
(or uncommon) our solar system should be, provided that it forms
according to the scenario of external radioactive enrichment.  We let
$d\pcl/dN$ be the probability density for a star forming within a
cluster containing $N$ members.  For the high end of the distribution
of cluster sizes $N$, we use the result that only about ten percent of
stars form within ``big'' clusters (e.g., see Roberts 1957; Elmegreen
and Clemens 1985; Battinelli and Capuzzo-Dolcetta 1991; Adams and Myers
2000; and others).  Here, ``big'' clusters are those sampled by the
observational cluster surveys, which are complete down to some minimum
cluster size $N_\star$ that is not precisely specified, but lies in
the range $100 < N_\star < 500$.  Since the putative birth cluster for
the solar system must be much larger, with $N \approx 2000$, it lies
safely in the size range for which the observational surveys are
complete.  Furthermore, the mass distribution for true (relatively
large) clusters can be represented by a power-law function $df/dN \sim
N^{-2}$ (Elmegreen and Efremov 1997). To obtain the probability density
function $d\pcl/dN$ for a star forming in a cluster of a given size, we
must multiply by one power of $N$ to obtain 
\be 
{d \pcl \over dN} = N {df \over dN} = \acl N^{-1} \, . 
\ee
We specify the normalization constant $\acl \approx (40 \ln10)^{-1}$ 
by integrating over the allowed range of cluster sizes (which we take 
to be $N_1 \equiv 10^2 \le N \le 10^6 \equiv N_2$). The probability 
${\cal P}$ that star forming environments provide the proper conditions 
for solar system formation is thus given by the integral 
\be
{\cal P} = \int_{N_1}^{N_2} {d \pcl \over dN} \, 
\psn P_{\rm dis} \, dN \, \approx 0.0085 \, . 
\label{eq:probint} 
\ee
In other words, the considerations of this paper imply that 1 out of
120 solar systems in the galaxy form in a dense enough environment to
be radioactively enriched by a supernova and, at the same time, a
sufficiently diffuse environment to allow outer planetary orbits to
remain relatively unperturbed.

The probability $\cal P$ depends on the parameters of the problem, in
particular the size $N_C$ of a cluster required for the solar system
to have optical depth unity to scattering events and the probability
$p_C$ for the cluster to contain a sufficiently massive star. One may
wish to consider varying scales $M_C$ for the required massive star,
or variations in the IMF, both of which change the value of
$p_C$. Similarly, one may wish to consider different types of
disruption events, which have different cross sections $\cross$, and
hence lead to different values of $N_C$.  The probability of a given
type of cluster environment, as given by equation [\ref{eq:probint}],
can be evaluated analytically as a power series for the standard 
case of $\mu$ = 2, i.e., 
\be
{\cal P} = {\acl \over 2} \sum_{k=0}^\infty (-1)^k \, 
\alpha^{k+1} \, {\Gamma [(k+1)/2] \over \Gamma [k+2] } \, , 
\label{eq:pseries} 
\ee
where $\Gamma(x)$ is the gamma function and where we have defined
$\alpha \equiv N_C (-\ln p_C)$, which has a ``standard'' value of
$\alpha \approx 1.2$. Unfortunately, however, the series converges
only for $\alpha \le 2$.  Because the dependence of $\cal P$ on the
fundamental physical quantities is somewhat opaque in equation
[\ref{eq:pseries}], we numerically evaluate the result and plot the
resulting probability $\cal P$ as a function of both $N_C$ and $M_C$
in Figure 2. As the mass scale required for enrichment increases to
$M_C$ = 60 $M_\odot$, for example, $\cal P$ decreases to about 0.0025,
which corresponds to odds of only 1 part in 400. The solar system is
thus much less likely to have been radioactively enriched by being 
born within an environment containing a 60 $M_\odot$ star than one 
containing a 25 $M_\odot$ star. 

\subsection{Radiation Environment of the Early Solar System} 

Given the constraints on the solar birth aggregate derived above, we
can place corresponding constraints on the radiation fields that the
early solar system experienced. These radiation fields can play an
important role in removing gas from the early solar nebula and can
thereby strongly influence planet formation (Shu, Johnstone, and
Hollenbach 1993; St{\"o}rzer and Hollenbach 1999). If the solar system
did indeed form within a large enough birth aggregate to provide
external radioactive enrichment, then the ultraviolet (UV) flux
incident upon the outer solar system will be dominated by the
background radiation field of the cluster rather than by the 
intrinsic radiation field of the early Sun. We substantiate this 
claim below. 

To determine the UV radiation field provided by the cluster
environment, we first need to estimate the distance from the solar
system to the massive stars in the birth aggregate (time averaged over
the planet formation epoch). The crossing time $t_{\rm cross}$ is
short compared to both the dynamical relaxation time $t_R$ and the
planet formation time $t_P$, but $t_P < t_R$, so we consider the solar
system to be on a random orbit in a fixed cluster structure (see eq.
[\ref{eq:timeorder}]). Current observations and theoretical simulations
indicate that the most massive stars may form in the cluster centers
(e.g., Bonnell and Davies 1998), so we take the massive stars to lie at
the origin. We must then calculate the expectation value for the total
UV flux, averaged over a typical orbit. For a centrally condensed
cluster (like a King model), the speed is nearly constant over the
orbit. For each pass through the cluster, the solar system experiences
an average flux $\langle F \rangle$ given by 
\be
\langle F \rangle \approx {1 \over 2R} \int_{-R}^{R} {L \over 4 \pi} 
{ds \over b^2 + s^2} \approx {L \over 8 R b} \, , 
\label{eq:orbmean} 
\ee
where $b$ is the impact parameter (the distance of closest approach to
the center). In a collapse model of cluster formation, the stellar
orbits are nearly radial outside the core and nearly isotropic inside
(Adams 2000); we therefore expect the typical impact parameter $b$ to
be given by the core radius. In open clusters, the core radii are
about ten times smaller than the cluster sizes (Binney and Tremaine
1987), so we take $b \approx R/10$. The quantity $L$ is the total
luminosity of the massive stars in the cluster center.  Since we are
interested in the ionizing ultraviolet flux $F_{uv}$, we take $L =
L_{uv}$, which is determined by integrating over the stellar initial
mass function. As before, we let the IMF take the form $df/d \ln m \sim$
$m^{-\gamma}$ for high mass stars.  We also assume that the ultraviolet 
luminosity can be modeled as a simple power-law over the stellar 
mass range of interest, i.e., $L_{uv} (m) = L_C (m/m_1)^q$. We take
$m_1$ = $M_{SN}$ = 8 $M_\odot$, the minimum stellar mass for a
supernova, because smaller stars have little contribution to the
ultraviolet radiation background. We also need to impose an upper mass
scale, which we take to be $m_2 = 100 M_\odot$, to keep the integrals
finite. Fitting the UV fluxes of massive stars, we find $q \approx 3.14$ 
and $L_C \approx 4.2 \times 10^{46}$ sec$^{-1}$ (ionizing photons per 
second). The mean UV luminosity $\langle L_{uv} \rangle$ thus becomes  
\be
\langle L_{uv} \rangle = \int_{m_1}^{m_2} 
A_m \, m^{-(1+\gamma)} \, dm \, L_C \, (m/m_1)^q \, = 
A_m L_C m_1^{-q} \, {1 \over q - \gamma} \bigl( m_2^{q-\gamma} - 
m_1^{q-\gamma} \bigr) \, , 
\label{eq:massmean} 
\ee
where $A_m$ is the normalization constant of the stellar IMF. 
Normalizing $A_m$ to account for the total number of stars $N$ 
in the cluster, we find $A_m = \fsn N \gamma m_1^\gamma$. 

Combining equations [\ref{eq:orbmean}] and [\ref{eq:massmean}], we
obtain the total ionizing UV flux impinging upon on the solar system 
from the background cluster environment 
\be
\fuv = {\fsn N L_C \over 8 b R} {\gamma \over q - \gamma} 
\Bigl[ \bigl( m_2 / m_1 \bigr)^{q-\gamma} - 1 \Bigr] \, . 
\label{eq:uvflux} 
\ee 
Inserting numerical values, we can evaluate this result to find
\be
\fuv = 1.6 \times 10^{12} {\rm cm}^{-2} {\rm sec}^{-1} \, 
\Bigl( {N \over 2000} \Bigr) 
\Bigl( {R \over 1 {\rm pc} } \Bigr)^{-2} \, . 
\ee
For comparison, the ionizing UV luminosity of the early Sun cannot be 
larger than about $L_{uv} \approx 10^{41}$ (Gahm et al. 1979; see also 
the discussion given in Shu et al. 1993) and hence the corresponding 
UV flux is found to be 
\be 
F_{uv \odot} = 3.5 \times 10^{13} {\rm cm}^{-2} {\rm sec}^{-1} \, 
\Bigl( {\varpi \over 1 {\rm AU} } \Bigr)^{-2} \, ,
\ee
where $\varpi$ is the (cylindrical) radial coordinate centered on the
Sun. The background flux of the cluster exceeds the local UV flux of
the Sun for radial positions $\varpi > 4.7$ AU, which is close to the
current value for the semi-major axis of Jupiter's orbit.  As a result, 
almost the entire region of the solar nebula that participates in
giant planet formation is dominated by the ionizing UV flux from the 
cluster environment, rather than the Sun. 

We can also compare the total number of ionizing UV photons intercepted
by the solar nebula from both the Sun and the background cluster. The
disk is embedded in the UV radiation field and both sides will be
exposed; the disk thus receives UV photons from the cluster at a rate
$\Phi_{uv}$ = $2 \pi R_{\rm d}^2 \fuv \approx 2 \times 10^{42}$
sec$^{-1}$, where $R_{\rm d}$ $\approx$ 30 AU is the radial size of
the disk. The disk will also intercept a fraction of the $\Phi_\odot$
$\approx 10^{41}$ sec$^{-1}$ UV photons generated by the nascent Sun.
For the limiting case of a flat disk which is optically thick but
spatially thin, the fraction of directly intercepted photons is 25\%
(Adams and Shu 1986); because the disk can be flared and because of
additional scattered (diffuse) photons, the actual fraction is
somewhat greater, about 50\% (Shu et al. 1993). The total rate of
intercepted solar UV photons is thus about $5 \times 10^{40}$
sec$^{-1}$. As a result, the cluster environment provides 40 times
more ionizing UV photons to the solar nebula than the Sun itself.

This enhanced UV flux can drive an enhanced rate of photoevaporation
from the disk. The mass loss rate for the simplest models (Shu et
al. 1993; Hollenbach et al. 1994) scale as ${\dot M}$ $\propto$
$\Phi_{uv}^{1/2}$, so the mass loss rate is (at least) $\sim$6.3 times
greater if the solar nebula lives within a large birth aggregate. For
some regimes of parameter space, the far-ultraviolet (as opposed to
ionizing UV) photons dominate the mass loss mechanism and the scalings
are different (St{\"o}rzer and Hollenbach 1999); one could perform an
analogous calculation for mass loss due to far-UV photons.  In any
case, for models of a disk immersed in the radiation field of a
cluster, the mass loss rate can be large ${\dot M}_D \approx 10^{-7}$
$M_\odot$ yr$^{-1}$ (again, see St{\"o}rzer and Hollenbach 1999). For
the minimum solar nebula with disk mass $M_D \approx 0.01 M_\odot$,
this mass loss rate would destroy the disk in only $10^5$ yr, far
shorter than the time scale required for giant planet formation 
(Lissauer 1993).  Although the details depend on the exact orbit of
the solar system through its birth environment and other undetermined
parameters, this putative cluster of $N=2000$ stars comes dangerously
close to preventing planet formation from taking place.

A related question is to ask what fraction of all stars are born in
sufficiently rich clusters so that the UV flux of the background
cluster dominates the intrinsic UV radiation field of the star. In
general, clusters large enough to be included in observational cluster
surveys (systems with a few hundred members or more) are large enough
to dominate the UV radiation field (Adams \& Myers 2000); as a result,
the fraction of stars that are exposed to intense radiation fields is
8--10 percent. About 90 percent of all solar systems thus have their
UV radiation fields dominated by their central stars (and can
presumably form planets without interference from the background 
environment).

\subsection{Effects on the Kuiper Belt} 

The large birth aggregate required to provide external radioactive
enrichment will also have a substantial impact on Kuiper Belt objects
(hereafter KBOs). The population of bodies in the Kuiper Belt is both
complex and still under investigation, but enough observations have
been made to provide preliminary constraints (e.g., see the reviews of
Jewitt and Luu 2000; Malhotra, Duncan, and Levison 2000; Farinella,
Davis, and Stern 2000; and references therein).  Briefly, the Kuiper
Belt contains a population of KBOs in nearly circular orbits with
semi-major axes in the range 30 -- 50 AU, a second population of KBOs
in resonances with Neptune, and a third population of KBOs with high
eccentricities and larger ($a>50$ AU) semi-major axes.  Although a
great deal of dynamical evolution has taken place between solar system
formation and the present-day observations, a (hypothetical) large
birth cluster will nonetheless have a dramatic impact on KBOs during
the first $\sim100$ Myr of solar system evolution.

To study the interplay between the Kuiper Belt and the solar birth
aggregate, the first step is to calculate the cross sections for the
scattering of KBOs by gravitational interactions with passing stars in
the birth cluster. The procedure is analogous to that described in \S
2.2. In this case, we start the scattering experiments with small
bodies in circular orbits. This suite of numerical experiments uses
orbital radii of $a$ = 30, 40, 50, 60, and 70 AU. Because KBOs are
small (the combined mass of the Kuiper Belt is estimated to be less
than an Earth mass), they act like test particles in the scattering
simulations. For computational convenience, we take all of the bodies
to have a mass of $10^{-6}$ $M_\odot$; this mass scale is small enough
that the bodies are indistinguishable from test particles and large
enough to allow the code to conserve energy and angular momentum to
good accuracy.  By including the KBO at $a$ = 30 AU (which would
clearly not survive because of Neptune), we obtain a consistency check
by comparing the results with those for Neptune scattering (Table 1).

The resulting cross sections for KBO scattering, listed here as a
function of the final (post-scattering) eccentricity, are shown in
Table 3.  Notice that the cross sections for the KBO at $a=30$ AU are
roughly comparable to those for Neptune (compare Tables 1 and 3); this
finding implies that even Neptune acts like a test body (for the most
part) during scattering interactions. We have not included the giant
planets in this set of scattering calculations. These planets
themselves can be scattered during the interactions and then can lead
to additional disruption of the putative KBOs; this secondary effect
is not included in this calculation and hence the cross sections
listed in Table 3 represent lower limits to possible disruption of KBO
orbits.

The scattering cross sections obtained here can be used in two ways:
We can assume that the solar birth aggregate is ``known'' (from the
previous results of this paper) and then predict undiscovered
properties of the outer Kuiper Belt. Alternately, we can use the
observed KBO populations to place further constraints on the solar
birth aggregate.

For example, if we assume that the previous sections specify the
properties of the solar birth cluster, then we know that it initially
contained $N \approx 2000$ members and the effective optical depth
$\tau$ to scattering events is given by 
\be 
\tau = {\cross \over (400 {\rm AU})^2} \bigl( {N \over N_C} \bigl)^2 
= {\cross \over (500 {\rm AU})^2} \, ,  
\ee 
where $\cross$ is the cross section for the scattering event of
interest. The corresponding probability of the given event {\it not}
occurring is thus $\exp[-\tau]$. Using the cross sections in Table 3,
we find that essentially all KBOs beyond $\sim50$ AU must attain
nonzero eccentricities. For example, KBOs at $a$ = 50 AU will
typically attain eccentricities $\epsilon > 0.2$ and KBOs at $a$ = 70
AU will attain $\epsilon > 0.4$. This type of dynamical excitation
(which also includes increased inclination angles of the orbits)
spreads out the KBO population and thereby lowers the apparent surface
density of objects on the sky; this reduction in surface density can
appear as an apparent ``edge'' to the solar system as recent
observations suggest (Allen, Bernstein, and Malhotra 2000). However, 
we must stress once again that the Kuiper Belt will undergo
substantial dynamical evolution of its own after the solar system
leaves its birth cluster.

Although the cross sections listed in Table 3 are relatively large, as
expected, it is significant that a large portion of the table has
cross sections less than the fiducial value $\cross \approx$ (400
AU)$^2$ required for solar system disruption (see \S 2.2 -- 2.3). As a
result, some fraction of the KBOs in the range 40 AU $\le a \le$ 70 AU
will survive the birth cluster. The KBOs can be removed (in the short
term) either through direct ejection (or capture) or by attaining a
large enough eccentricity to cross the orbit of Neptune.  Although
KBOs with larger radii ($a$) have larger cross sections, they need to
be scattered to larger eccentricities to encounter Neptune. These two
trends nearly compensate for each other and yield a nearly constant
cross section for KBOs to be (promptly) removed from the solar system:
$\cross \approx$ (350 AU)$^2$. This value implies that the probability
of KBO survival runs at about $\exp[-\tau] \approx$ 0.6.  In other
words, about 40 percent of the KBOs will be removed from the solar
system while the Sun remains in its birth cluster (and an additional
population will be removed later through longer term dynamical
interactions).

As an alternate approach, we can use our results from KBO scattering
to place further constraints on the solar birth aggregate. We first
note that we could use the survival of the Kuiper Belt as the
criterion for solar system to not be disrupted. We then repeat the
analysis of \S 2.3 -- 2.5 using the cross section for KBO removal
$\cross \approx$ (350 AU)$^2$. Because this cross section is somewhat
lower than that used previously (that for disruption of the giant
planet orbits), the derived constraints on the solar birth cluster are
correspondingly weaker: $\nexp$ = 2250 $\pm$ 1250 and $\cal P$ =
0.0095. Using only the survival of the Kuiper Belt, we thus obtain a
(1 $\sigma$) upper limit on the size of the birth cluster: 
$N \le 3500$. 

We can also repeat the probability analysis by requiring that the 
solar system survive with both its giant planet orbits {\it and} 
its Kuiper Belt intact.  In this case, the joint probability 
distribution $P_\odot$ for solar system survival takes the form 
\be
P_\odot (N) = \psn P_{\rm dis} = 
\bigl (1 - p_C^N \bigr) \, \exp \bigl[ - (N/N_C)^\mu \bigr] \, 
\exp \bigl[ - (N/N_{CK})^\mu \bigr] \, , 
\label{eq:joint2} 
\ee 
where the second exponential factor represents the survival of the
Kuiper Belt (and where we consider the two processes to be independent).  
The quantity $N_{CK}$ $\approx$ 2860 is the number of stars in the
cluster required to make the scattering optical depth unity for
scattering interactions leading to KBO removal.  In this case, because
we have an added constraint, the resulting bounds are somewhat more
restrictive than before: $\nexp$ = 1520 $\pm$ 827 and $\cal P$ =
0.0067 (which corresponds to odds of one part in 150).

Before leaving this section, we note that the Oort cloud of comets 
may be even easier to disrupt than the Kuiper Belt. For a given model 
of comet formation, one could thus find the corresponding constraints 
on the birth aggregate of the solar system. We leave this issue for 
future work.  

\section{CONCLUSIONS and DISCUSSION} 

In this paper, we have explored the consequences of the solar system
being formed within a group environment that is large enough to
contain a massive star that enriches the early solar system in
radioactive species and is also sufficiently diffuse to allow the
planetary orbits to remain unperturbed. In particular, we have
obtained the following results:

[1] We have calculated the cross sections for the outer planets in 
our solar system to experience orbital changes due to scattering
interactions with binary systems in a cluster environment. We find 
the cross sections for eccentricity increases (see Table 1) and for 
increases in the relative inclination angles of the planetary orbits
(see Table 2). The cross section for the scattering events to increase
either the eccentricities or the inclination angles beyond the 
currently observed values is $\cross \approx$ (400 AU)$^2$. The cross
section for planetary ejection and/or capture is somewhat lower, about
$\cross \approx$ (130 AU)$^2$. However, all of these cross sections
are substantially larger than the area subtended by the solar system,
{\sl Area} $\approx \pi$ (30 AU)$^2$. 

[2] We have estimated the probability distribution for the number $N$
of stars in the birth aggregate for our solar system (see Figure 1).
Using the coupled constraints that the group was large enough to
contain a massive star (to enrich the solar system in radioactive
elements) and small enough so that the outer planetary orbits are not
severely disrupted, we find that $N = \nexp \approx 2000 \pm 1100$.
The expectation value $\nexp$ varies relatively slowly with the
parameters of the problem and has the analytic solution given by
equation [\ref{eq:ngeneral}]. 

[3] The {\it a priori} probability for a star being born in the type
of environment required for the external enrichment scenario for our
solar system (i.e., subject to the probability distribution depicted
in Figure 1) is ${\cal P} \approx 0.0085$.  The odds of the solar
system forming in this type of environment are thus about 1 in 120.
This result can be readily generalized to accommodate other choices 
of parameters (see eq. [\ref{eq:pseries}] and Figure 2). 

[4] The time scales for the cluster environment obey a particular
ordering (see equation [\ref{eq:timeorder}]): The crossing time
$t_{\rm cross}$ of the cluster is much shorter than the time $t_P$
required to form giant planets, so the solar system randomly samples
the cluster environment over that epoch. The dynamical relaxation time
$t_R$ is somewhat longer than $t_P$, and the total cluster lifetime
$\tcl \gg t_P$, so the cluster does not change its structure
appreciably while the planets form. 

[5] Within the scope of this external enrichment scenario, we have
reconstructed the radiation field provided by the birth aggregate of
the solar system.  The early solar nebula receives about 40 times more
ionizing UV photons from the background cluster environment than from
the early Sun. Only the inner portion of the nebula, at radii $r < 5$
AU, has its ionizing UV flux dominated by the Sun.  This intense flux
of UV radiation can severely ablate the early solar nebula and greatly 
compromise giant planet formation.

[6] Objects forming in the Kuiper Belt will also be scattered due to
interactions in the birth cluster. We have calculated the scattering
cross sections for KBOs on initially circular orbits with radii in the
range 30 AU $\le a \le$ 70 AU (Table 3). The cross sections for prompt
removal are $\cross \approx$ (350 AU)$^2$ over the outer part of this
radial range. If we include the required survival of the Kuiper Belt
in the probability analysis, we obtain slightly tighter constraints on
the solar birth aggregate: $\nexp$ = 1520 $\pm$ 827, $\cal P$ =
0.0067, and {\it a priori} odds of one part in 150.

Some authors have suggested that the formation of our solar system
must be triggered by the same supernova that is postulated to provide
the radioactive enrichment (e.g., Boss and Foster 1998). We stress here
that external enrichment does not necessarily imply a triggered
collapse.  The time scale for cluster formation is relatively short (a
few Myr; e.g., Elmegreen 2000) and the time scale for the collapse of
an individual star forming site is much shorter (about 10$^5$ yr;
e.g., Myers and Fuller 1993, Adams and Fatuzzo 1996). Thus, the solar
system could be formed within a cluster and yet be formed through a
spontaneous (un-triggered) collapse. Another motivation for the
collapse being triggered is that the ambipolar diffusion time scale is
generally long (about $10^7$ years), too long for the survival of the
necessary radioactive nuclei. However, observational evidence shows
that the time scale for molecular cloud cores to shed their magnetic
support and begin dynamic collapse is much shorter (about 1 Myr) for
all cores (e.g., Jijina et al. 1999; Myers and Lazarian 1998); the
ambipolar diffusion time scale is thus not an insurmountable obstacle
for star formation in clusters.

The results of this paper have important ramifications for the ongoing
debate concerning radioactive enrichment of the early solar system.
The meteoritic data strongly indicate that such enrichment took place.
However, both the external scenario of enrichment by a massive star
and the internal scenario of self-enrichment have some difficulty
reproducing all of the short-lived radioactive species (e.g., see
Goswami and Vanhala 2000; Lee et al. 1998; and references therein).
Although the results are not definitive, this paper tends to favor 
the self-enrichment scenario for two reasons: (1) A star forming 
environment that is simultaneously large enough to provide external
enrichment and diffuse enough to not disrupt the planetary orbits is
{\it a priori} an unlikely event (at the level of 1 part in 120). (2)
If the solar system formed in the large birth aggregate required for
external enrichment, the corresponding radiation fields are likely to
have compromised giant planet formation.

Nevertheless, the external enrichment scenario is not conclusively
ruled out. Events with long odds (100:1) do indeed sometimes happen.
If one adopts the external enrichment scenario, a consistent solution
exists and the results of this paper place tight constraints on the
birth environment of our solar system: The birth cluster contained $N
\approx 2000$ other stars and subjected the early solar nebula to an
intense UV radiation field. Scattering interactions in the birth
cluster would then be responsible (at least in part) for the observed
nonzero (but small) inclination angles and eccentricities of the giant
planet orbits.  Furthermore, objects in the Kuiper Belt would be
dynamically excited by these scattering interactions. This highly
disruptive environment must be accounted for in a complete description
of solar system formation.

\bigskip 

During the preparation of this paper, we have greatly benefited from
discussions with Gary Bernstein, Pat Cassen, Gus Evrard, Dave
Hollenbach, Nathan Schwadron, Frank Shu, and Kevin Zahnle. This work
was supported by NASA Grant No. 5-2869 and the University of Michigan.

\newpage 

\bigskip 
\centerline{}
\bigskip 
 
\centerline{\bf Table 1} 
\medskip 
\centerline{\bf Cross Sections for Planet Scattering: Eccentricity Increase}
\bigskip 
\centerline{[all cross sections in units of (AU)$^2$]} 
\bigskip 
 
\begin{center}
\begin{tabular}{ccccc}
\hline 
\hline
\hline 
$\epsilon$ & Jupiter & Saturn & Uranus & Neptune \\
\hline 
\hline 
0.05& 54300 $\pm$ 707 & 67700 $\pm$ 775&126000 $\pm$ 1030 & 167000 $\pm$ 1150\\
0.10& 40700 $\pm$ 611 & 55000 $\pm$ 700&106000 $\pm$  958 & 143000 $\pm$ 1080\\
0.20& 29200 $\pm$ 521 & 42900 $\pm$ 621& 83300 $\pm$  860 & 113000 $\pm$ 983\\
0.30& 22800 $\pm$ 458 & 36700 $\pm$ 576& 69500 $\pm$  794 & 95600 $\pm$  915\\
0.40& 18600 $\pm$ 410 & 32400 $\pm$ 544& 59300 $\pm$  738 & 82900 $\pm$  861\\
0.50& 15500 $\pm$ 374 & 28800 $\pm$ 517& 52300 $\pm$  697 & 73300 $\pm$  816\\
0.60& 13700 $\pm$ 351 & 25800 $\pm$ 491& 46600 $\pm$  661 & 64900 $\pm$  774\\
0.70& 11900 $\pm$ 327 & 22700 $\pm$ 462& 41500 $\pm$  626 & 58700 $\pm$  740\\
0.80& 10500 $\pm$ 306 & 19800 $\pm$ 432& 36700 $\pm$  592 & 53200 $\pm$  710\\
0.90&  9120 $\pm$ 286 & 17500 $\pm$ 407& 32300 $\pm$  558 & 47700 $\pm$  678\\
0.95&  8520 $\pm$ 276 & 16400 $\pm$ 393& 30400 $\pm$  542 & 45100 $\pm$  662\\
1.00&  7970 $\pm$ 267 & 15300 $\pm$ 381& 28400 $\pm$  526 & 41900 $\pm$  640\\
\hline 
\hline 
escape& 7290 $\pm$ 255 &14000 $\pm$ 365 &25000 $\pm$ 492 &35400 $\pm$ 584\\
capture& 684 $\pm$ 81.4 &1300 $\pm$ 115 & 3320 $\pm$ 194 & 6640 $\pm$ 280\\
\hline
\hline
\hline 
\end{tabular}
\end{center}

\newpage 
\bigskip 
\centerline{}
\bigskip 
 
\centerline{\bf Table 2} 
\medskip 
\centerline{\bf Cross Sections for Planet Scattering: Inclination Increase} 
\bigskip 
\centerline{[all cross sections in units of (AU)$^2$]} 
\bigskip 
 
\begin{center}
\begin{tabular}{cccc}
\hline 
\hline
\hline 
$\Delta \theta_i$ & angle increase & total \\ 
\hline 
\hline 
0.06 & 140900 $\pm$1100 & 158000 $\pm$1130\\
0.10 & 114600 $\pm$1010 & 132000 $\pm$1050\\
0.20 &  85000 $\pm$ 880 & 102000 $\pm$ 943\\
0.30 &  71800 $\pm$ 815 &  89000 $\pm$ 888\\
0.40 &  63600 $\pm$ 770 &  80800 $\pm$ 850\\
0.50 &  57600 $\pm$ 736 &  74800 $\pm$ 822\\
0.60 &  53100 $\pm$ 707 &  70300 $\pm$ 798\\
0.70 &  49500 $\pm$ 684 &  66700 $\pm$ 780\\
0.80 &  46400 $\pm$ 663 &  63600 $\pm$ 763\\
0.90 &  43900 $\pm$ 647 &  61000 $\pm$ 750\\
1.00 &  41400 $\pm$ 629 &  58500 $\pm$ 736\\
1.10 &  39500 $\pm$ 616 &  56600 $\pm$ 726\\
1.20 &  37700 $\pm$ 602 &  54800 $\pm$ 715\\
1.30 &  35700 $\pm$ 585 &  52800 $\pm$ 702\\
1.40 &  33700 $\pm$ 570 &  50900 $\pm$ 690\\
1.50 &  32400 $\pm$ 559 &  49500 $\pm$ 682\\
1.60 &  30900 $\pm$ 546 &  48000 $\pm$ 672\\
1.70 &  29400 $\pm$ 534 &  46500 $\pm$ 663\\
1.80 &  27700 $\pm$ 518 &  44900 $\pm$ 652\\
2.00 &  24600 $\pm$ 488 &  41700 $\pm$ 630\\
2.20 &  21400 $\pm$ 459 &  38500 $\pm$ 609\\
2.40 &  17100 $\pm$ 411 &  34300 $\pm$ 576\\
2.60 &  12300 $\pm$ 349 &  29400 $\pm$ 537\\
2.80 &   6870 $\pm$ 263 &  24000 $\pm$ 490\\
3.00 &   2020 $\pm$ 143 &  19100 $\pm$ 441\\
\hline 
\hline 
\hline 
\end{tabular}
\end{center}

\newpage 
\bigskip 
\centerline{}
\bigskip 
 
\centerline{\bf Table 3} 
\medskip 
\centerline{\bf Cross Sections for KBO Scattering} 
\bigskip 
\centerline{[all cross sections in units of (AU)$^2$]} 
\bigskip 
 
\begin{center}
\begin{tabular}{cccccc}
\hline 
\hline
\hline 
$\epsilon$ & $a$ = 30 AU & $a$ = 40 AU & $a$ = 50 AU & $a$ = 60 AU & $a$ = 70 AU \\
\hline 
\hline 
0.05 &172000 $\pm$ 1670 &206000 $\pm$ 1780 &236000 $\pm$ 1860 &263000 $\pm$ 1780 & 286000 $\pm$ 1820\\ 
0.10 &147000 $\pm$ 1580 &179000 $\pm$ 1700 &206000 $\pm$ 1780 &229000 $\pm$ 1720 & 251000 $\pm$ 1770\\
0.20 &119000 $\pm$ 1450 &145000 $\pm$ 1570 &169000 $\pm$ 1670 &189000 $\pm$ 1630 & 209000 $\pm$ 1680\\
0.30 & 99600 $\pm$ 1350 &123000 $\pm$ 1480 &143000 $\pm$ 1570 &161000 $\pm$ 1540 & 179000 $\pm$ 1600\\
0.40 & 86700 $\pm$ 1270 &106000 $\pm$ 1390 &126000 $\pm$ 1500 &140000 $\pm$ 1460 & 156000 $\pm$ 1530\\
0.50 & 76100 $\pm$ 1200 & 92800 $\pm$ 1310 &110000 $\pm$ 1420 &123000 $\pm$ 1390 & 138000 $\pm$ 1460\\
0.60 & 66700 $\pm$ 1120 & 82900 $\pm$ 1250 & 98600 $\pm$ 1350 &109000 $\pm$ 1320 & 123000 $\pm$ 1400\\
0.70 & 59600 $\pm$ 1070 & 74100 $\pm$ 1190 & 88300 $\pm$ 1300 & 97800 $\pm$ 1270 & 111000 $\pm$ 1350\\
0.80 & 52000 $\pm$ 1000 & 66400 $\pm$ 1130 & 78800 $\pm$ 1240 & 87300 $\pm$ 1210 &  99400 $\pm$ 1290\\
0.90 & 46000 $\pm$  947 & 59100 $\pm$ 1080 & 70400 $\pm$ 1180 & 77800 $\pm$ 1160 &  88900 $\pm$ 1230\\
1.00 & 39800 $\pm$  889 & 52700 $\pm$ 1030 & 61800 $\pm$ 1110 & 67700 $\pm$ 1090 &  78000 $\pm$ 1170\\ 
\hline 
\hline 
\hline 
\end{tabular}
\end{center}

\newpage 
\centerline{\bf FIGURE CAPTIONS} 
 
\figcaption[] 
{Probability distributions for the number of stars in the solar birth
aggregate. The dashed curves show the probability $\psn(N)$ of a
cluster containing a sufficiently massive star for radioactive
enrichment (the increasing function of $N$) and the probability
$P_{\rm dis}(N)$ of the solar system remaining undisrupted (the
decreasing function of $N$). The solid curve shows the joint
probability $P_\odot(N)$ of the solar birth aggregate being 
simultaneously large enough to contain a massive star (with mass
$M_\ast > M_C$ = 25 $M_\odot$) and small enough to allow the 
planetary orbits to not be disrupted.} 

\figcaption[]
{Probability of the solar system being born in a cluster environment 
that is rich enough to contain a massive star (with mass greater than 
$M_C$) and diffuse enough to not disrupt the orbits of the giant planets. 
The solid line shows the probability $\cal P$ as a function of the 
number $N_C$ of stars required to make the birth aggregate optically 
thick to scattering events. The dashed curve shows the probability 
$\cal P$ as a function of the required mass scale $M_C$ (where the 
numbers on the horizontal axis must be divided by 100 to express 
the mass $M_C$ in units of $M_\odot$). }

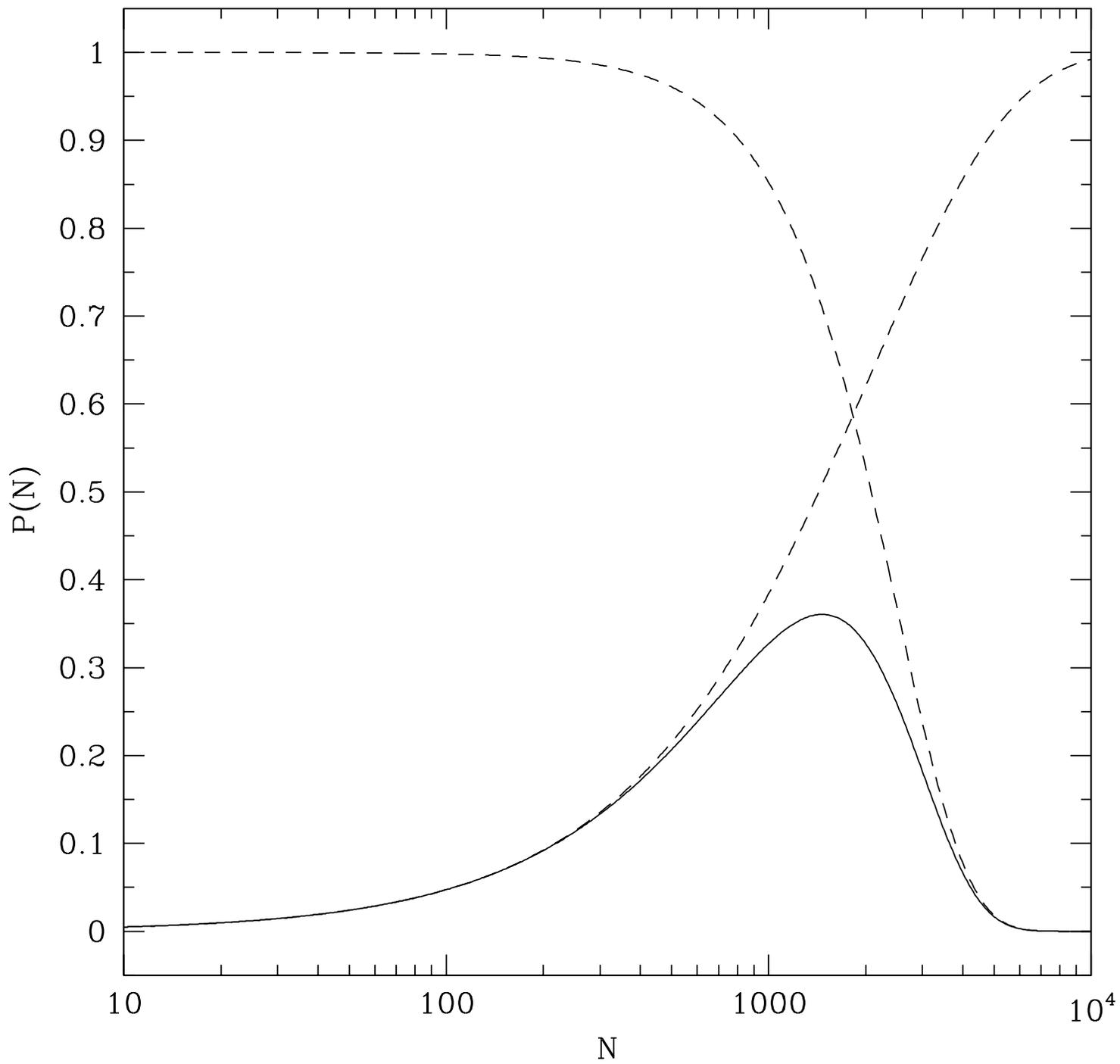
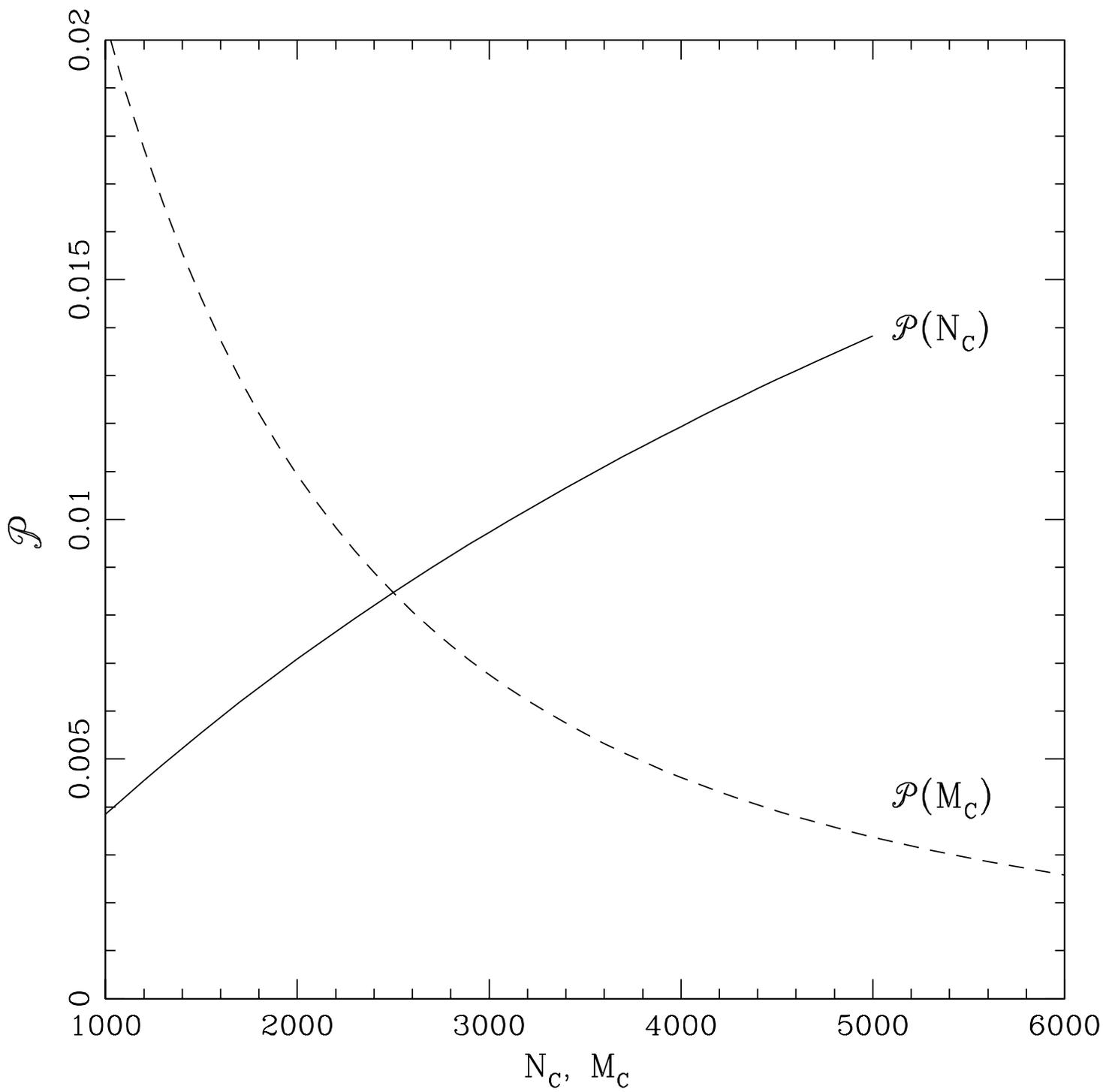
\end{document}